\documentclass[sigconf]{acmart}
\pdfoutput=1
\usepackage{booktabs} 
\usepackage{amsmath,bm}
\usepackage{amsthm, amssymb}
\usepackage[linesnumbered,ruled,vlined]{algorithm2e}
\usepackage{graphicx}
\usepackage{caption}
\usepackage{subcaption}
\usepackage{pbox}
\usepackage{romannum}
\usepackage{enumitem}
\usepackage{soul}
\usepackage{pdfpages}
\usepackage{flushend}
\usepackage{bbm}

\DeclareMathOperator*{\argmin}{arg\,min}

\newcommand{\Thetabold}{\ensuremath{\mathbf \Theta}}
\newcommand{\R}{\ensuremath{\mathbb R}}
\newcommand{\matV}{\ensuremath{\mathbf V}}
\newcommand{\matU}{\ensuremath{\mathbf U}}

\newcommand{\matX}{\ensuremath{\mathbf X}}
\newcommand{\matA}{\ensuremath{\mathbf A}}
\newcommand{\matB}{\ensuremath{\mathbf B}}
\newcommand{\matG}{\ensuremath{\mathbf G}}
\newcommand{\matI}{\ensuremath{\mathbf I}}
\newcommand{\matS}{\ensuremath{\mathbf S}}

\newcommand{\vecv}{\ensuremath{\mathbf v}}
\newcommand{\vecu}{\ensuremath{\mathbf u}}

\newcommand{\vecx}{\ensuremath{\mathbf x}}
\newcommand{\vecg}{\ensuremath{\mathbf g}}
\newcommand{\veca}{\ensuremath{\mathbf a}}
\newcommand{\vecb}{\ensuremath{\mathbf b}}

\newcommand{\vecd}{\ensuremath{\mathbf d}}

\newcommand{\spara}[1]{\smallskip\noindent{\bf{#1}}}

\newtheorem{problem}{Problem}

\usepackage[show]{chato-notes}

\makeindex
\begin{document}
\title[Fighting Fire with Fire]{Fighting Fire with Fire: Using Antidote Data to Improve Polarization and Fairness of Recommender Systems}


\author{Bashir Rastegarpanah}
\affiliation{\institution{Boston University}}
\email{bashir@bu.edu}

\author{Krishna P. Gummadi}
\affiliation{\institution{MPI-SWS}}
\email{gummadi@mpi-sws.org}

\author{Mark Crovella}
\affiliation{\institution{Boston University}}
\email{crovella@bu.edu}


\begin{abstract}
  The increasing role of recommender systems in many aspects of
  society makes it essential to consider how such systems may impact
  social good.  Various modifications to recommendation algorithms have
  been proposed to improve their performance for specific socially
  relevant measures.  However, previous proposals are often not easily
  adapted to different measures, and they generally require the ability
  to modify either existing system inputs, the system's algorithm, or
  the system's outputs. As an alternative, in this paper we introduce the
  idea of improving the social desirability of recommender system
  outputs by adding more data to the input, an approach we view as
  providing `antidote' data to the system. We formalize the antidote
  data problem, and develop optimization-based solutions. We take as our
  model system the matrix factorization approach to recommendation, and
  we propose a set of measures to capture the polarization or fairness
  of recommendations.  We then show how to generate antidote data for
  each measure, pointing out a number of computational efficiencies, and
  discuss the impact on overall system accuracy.
  Our experiments show that a modest budget for antidote data can lead
  to significant improvements in the polarization or fairness of
  recommendations.
\end{abstract}

%
%
%


\copyrightyear{2019}
\acmYear{2019}
\setcopyright{acmlicensed}
\acmConference[WSDM '19]{The Twelfth ACM International Conference on Web Search and Data Mining}{February 11--15, 2019}{Melbourne, VIC, Australia}
\acmBooktitle{The Twelfth ACM International Conference on Web Search and Data Mining (WSDM '19), February 11--15, 2019, Melbourne, VIC, Australia}
\acmPrice{15.00}
\acmDOI{10.1145/3289600.3291002}
\acmISBN{978-1-4503-5940-5/19/02}

\fancyhead{}

\maketitle

\section{Introduction}
Recommender systems are at the core of many online platforms that
influence the choices we make in our daily lives ranging from what
news we read (e.g., Facebook, Twitter) and whose products and services
we buy (e.g., Amazon, Uber, Netflix) to whom we meet (e.g., OKCupid,
Tinder).  As users increasingly rely on recommender systems to make
life-affecting choices, concerns are being raised about their
inadvertent potential for social harm. Recently, studies have shown
how recommender systems predicting user preferences might offer {\it
unfair or unequal} quality of service to indvidual (or groups of)
users~\cite{beutel2017beyond, burke2018balanced} or lead to {\it
societal polarization} by increasing the divergence between
preferences of individual (or groups of)
users~\cite{Dandekar09042013}.

\sloppypar
Collaborative filtering recommender systems rely on user-provided data
to learn models that are used to predict unknown user preferences. As
a result, the recommendations made by such systems may carry undesired
properties which are inherent in the observed data.  A natural
approach then is to consider transformations of input data that
ameliorate those properties.

In this paper we explore a new approach.  Rather than transforming the
system's exisiting input data, we investigate whether
simply \emph{augmenting the input with additional data} can improve
the social desirability of the resulting recommendations.  We explore
this question by developing a generic framework that can be used to
improve a variety of socially relevant properties of recommender
systems.  Our framework turns a technique that has previously been
thought of as anti-social attacks on learning systems into a method
with socially desirable outcomes.

As a strategy for improving recommendations, the data augmentation
approach has multiple advantages.  Adding new input data may be easier
than modifying existing data inputs, as when a system is already
running.  Additional data can be provided to the system by a
third-party who does not need the ability to modify the system's
existing input, nor the ability to modify the system's algorithms.
Further, the approach is applicable to a wide range of socially
relevant properties of a system -- essentially any property that can
be expressed as a differentiable function of the systems inputs
(ratings) and/or outputs (predictions).

\sloppypar
The framework we develop starts from an existing matrix-factorization
recommender system organized according to users and items, that has
already been trained with some input (ratings) data.  We consider the
addition to the system of new users who provide ratings of existing
items.  The new users' ratings are chosen according to our framework,
so as to improve a socially relevant property of the recommendations
that are provided to the \emph{original} users.  We call the
additional ratings provided `antidote' data (by analogy to existing
work studying data poisoning).

In this paper we instantiate the framework by proposing metrics that
capture the polarization and unfairness of the system's
recommendations.  These metrics build on and extend previous
proposals, and include measures of both individual and group
unfairness.  We show how to generate antidote data for these metrics,
and we present a number of computational efficiencies that can be
exploited.  In the process we consider the relationship between
improvements to socially-relevant measures and changes to overall
system accuracy. Finally, we show that the small amounts of antidote
data (typically on the order of 1\% new users) can generate a dramatic
improvement (on the order of 50\%) in the polarization or the fairness
of the system's recommendations.

\section{Related Work}
\label{sec:relwork}
In this section, we first discuss how our measures of fairness and
polarization for recommender systems relate to those discussed in
prior works. Later, we describe how we leverage insights and methods
explored in adversarial machine learning to cause social harm towards
social good in recommender systems.


\textbf{Fairness in machine learning and recommender systems:} The
past years have witnessed a growing awareness about the potential for
social harm by the use of machine learning algorithms in
life-affecting decision making
scenarios~\cite{barocas2016big, boyd2012critical}. In response, researchers have
proposed numerous notions and measures of fairness for machine
learning tasks as varied as classification~\cite{zafar2017training,
  zafar2017fairness, hardt2016equality, zemel2013learning},
regression~\cite{berk2017convex},
ranking~\cite{biega2018equity, singh2018fairness, zehlike2017fa}, and
set selection~\cite{celis2016fair}. These proposed notions
fall under two broad categoies: those measuring unfairness at the
level of {\it individual users} and those that measure unfairness at
the level of {\it user groups}~\cite{dwork2012fairness}. The
group-level unfairness measures can be further sub-divided into those
that prohibit the use of information related to a user's sensitive
group membership when making predictions and those that require users
belonging to different sensitive groups to receive, on average, {\it
  equal quality of service}. The quality of service received by a user
can in turn be measured either {\it conditioned or unconditioned} on
the service outcomes deserved by the user.

Compared to learning tasks such as classification and regression, few
studies have explored fairness notions in the context of recommender
systems. Recently, Burke et al. \cite{burke2018balanced} observed that
recommender systems predicting user preferences over items would have
to consider fairness from {\it two-sides} namely, from the perspective
of {\it users} receiving the recommendations and from the perspective
of {\it items} being recommended. Some of the early works by Kamishima
et. al. \cite{kamishima2012enhancement, pmlr-v81-kamishima18a,
  kamishima2017considerations} focussed on notions of group-level
fairness, where the learning model is modified to ensure that item
recommendations are independent of users' features revealing sensitive
group membership such as race and gender. More recently, Beutel
et. al. \cite{beutel2017beyond} and Yao et. al. \cite{yao2017beyond}
have defined notions of group-level fairness in recommender systems
based on the accuracy of predictions across different groupings of
users or items.

\textbf{Novel Contributions:} Here, we not only build upon the
group-level notions of fariness proposed by Beutel and Yao (by
generalizing them to scenarios with more than two groups), but we also
extend them to individual-level. We further note that our fairness
notions can be applied either from the perspective of users or items.

\textbf{Mechanims for fair machine learning and recommender systems:}
Prior works have explored a number of approaches to incorporating
fairness in learning models and recommender systems. These approaches
can be broadly categorized into those that rely on (i) {\it
  pre-processing}, i.e., transform the training data to reduce the
potential for unfair outcomes when using traditional learning
models~\cite{kamiran2012data,calmon2017optimized}, (ii) {\it in-processing}, i.e., change the learning
objectives and models to ensure fair outcomes even using unmodified
training data~\cite{kamishima2011fairness,agarwal2018reductions}, and (iii) {\it post-processing}, i.e., modify
potentially unfair outcomes from existing pre-trained learning
models~\cite{hardt2016equality, corbett2017algorithmic}.  

\textbf{Novel Contributions:} In this paper, we explore a
different approach to incorporating our fairness notions in
recommender systems. Our approach is in contrast to existing
approaches to fair recommendations that primarily rely on
in-processing~\cite{burke2018balanced, kamishima2012enhancement}. Unlike in-processing approaches, our approach does not require us to modify the recommendation algorithm for each of our desired notions of fairness.

\textbf{Leveraging adversarial machine learning for social good:} Our
approach relies on methods that have been traditionally used in
adversarial learning literature to cause social
harm \cite{huang2011adversarial}. Our key insight is that we can
retarget adversarial methods designed to ``poison'' training data and
cause social harm to generate ``antidote'' training data for social
good. Specifically, our antidote data generation methods are inspired
by prior work on data poisoning attacks on factorization-based
collaborative filtering \cite{li2016data}. 

Most pre-processing approaches target learning new fair (latent and
transformed) representations of original data. Recently, Beutel
et. al. \cite{beutel2017data} leveraged
adversarial training procedures to remove information about sensitive
group membership from the latent representations learned by a neural
network. In contrast, our approach leaves the original training data
untouched and instead adds new antidote data to achieve 
fairness objectives. As our evaluation results presented later in the
paper will show, by leaving the original training data unmodified, our
approach also achieves good overall prediction accuracy (the
traditional objective of recommender algorithms).

\textbf{Polarization:} Polarization refers to the degree to which
opinions, views, and sentiments diverge within a population. Several prior works have raised and explored concerns that
recommender systems might increase societal polarization by tailoring
recommendations to individual user's preferences and trapping users in
their own ``personalized filter bubbles''~\cite{pariser2011filter, hannak2013measuring}. Dandekar
et. al~\cite{Dandekar09042013} show how many traditional
recommender algorithms used on Internet platforms can lead to
polarization of user opinions in society. 

\textbf{Novel Contributions:} We propose to measure the
  polarization of a recommender system as the extent to which
  predicted ratings for items vary (diverge) across users. Our polarization metric is consistent with those proposed in~\cite{Dandekar09042013, matakos2017measuring}.
  We show how our antidote data generation framework can be used to target reducing (or in certain scenarios, increasing) polarization in predicted ratings.
\section{Optimal Antidote Data Problem}
We start by presenting the system setup, notation, and problem definition.
Assume $\matX \in \R^{n \times d}$ is a partially observed rating matrix of $n$ users and $d$ items such that element $x_{ij}$ denotes the rating given by user $i$ to item $j$. Let $\Omega$ be the set of indices of known ratings in $\matX$. Also $\Omega^i$ denotes the indices of known item ratings for user $i$, and $\Omega_j$ denotes the indices of known user ratings for item $j$.

For a matrix $\matA$, $P_\Omega(\matA)$ is a matrix whose elements at $(i,j) \in \Omega$ are $a_{ij}$ and zero elsewhere. Similarly, for a vector $\veca$, $P_{\Omega_j}(\veca)$ is a vector whose elements at $i \in \Omega_j$ are the corresponding elements of $\veca$ and zero elsewhere. Throughout the paper, we denote the column $j$ of $\matA$ by the vector $\veca_j$ and the row $i$ of $\matA$ by the vector $\veca^i$. All vectors are column vectors.

We assume a factorization based collaborative filtering algorithm is applied to estimate the unknown ratings in $\matX$, i.e., for each user $i$ and item $j$ we find $\ell$-dimensional representations $\vecu_i$ and $\vecv_j$ such that $\ell<<min(n,d)$ and the rating $x_{ij}$ is modeled by $x_{ij} \approx \vecu_i^\intercal \vecv_j$. 

More specifically, we consider a factorization algorithm $\Thetabold$ that finds factors $\matU \in \R^{\ell \times n}$ and $\matV \in \R^{\ell \times d}$ by solving the following optimization problem:
\begin{equation}
\label{eq:MFobj}
\underset{\matU,\matV}{\argmin} \quad ||P_\Omega (\matX - \matU^\intercal \matV)||_F^2 + \lambda (||\matU||_F^2 + ||\matV||_F^2)
\end{equation}
where columns of $\matU$ are the user latent vectors, and columns of $\matV$ are the item latent vectors. The first term in \ref{eq:MFobj} denotes the estimation error over known elements of $\matX$ and the second term is an $\ell_2$-norm regularizer added to avoid overfitting. The unknown ratings are then estimated by setting $\hat{\matX} = \matU^\intercal \matV$.

\begin{figure}[t]
    \centering
    \captionsetup{width=0.5\textwidth}
    \includegraphics[width=0.5\textwidth]{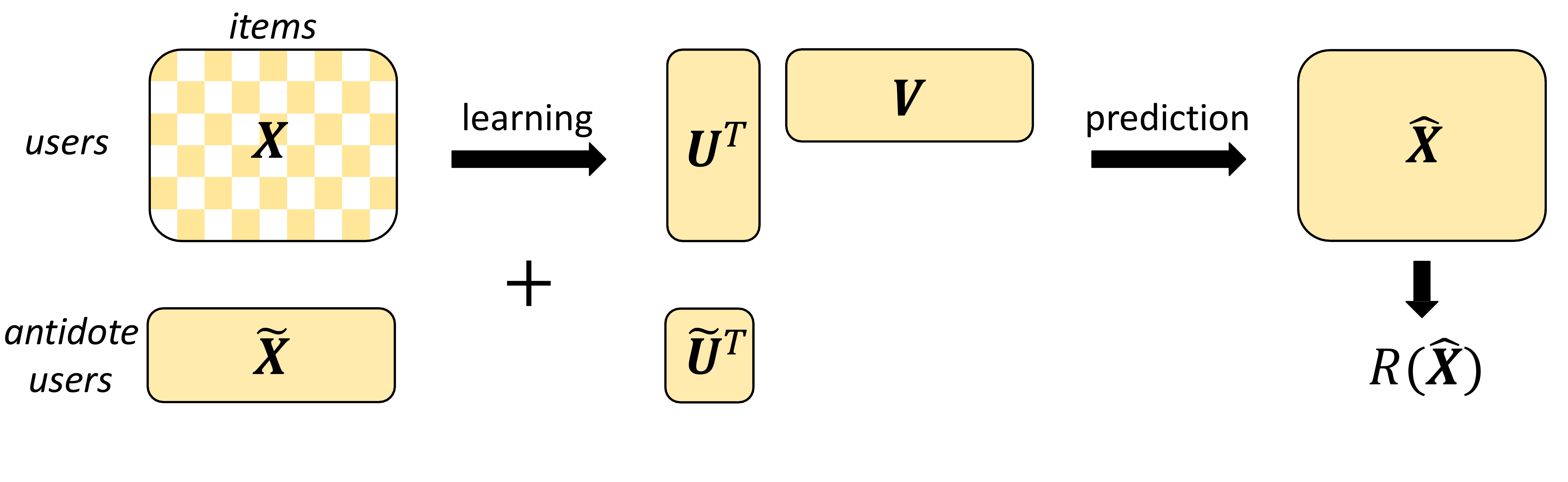}
    \caption{\footnotesize The effect of antidote data on a matrix factorization system. Initially the system learns factors $\matU$ and $\matV$ from a partially observed rating matrix $\matX$. The latent factors are then used to find the estimated rating matrix $\hat{\matX}$ which is an input to the socially relevant metric $R$. Adding antidote ratings $\tilde{\matX}$ introduces the new user latent factor $\tilde{\matU}$ and modifies the item latent factor $\matV$, generating a new $\hat{\matX}$ that improves $R(\hat{\matX})$.}\label{fig:model}
\vspace*{-\baselineskip}
\end{figure}

We can think of our factorization algorithm as a function that maps a partially observed rating matrix $\matX$ to matrices $\matU$ and $\matV$, and has additional parameters $\ell$ and $\lambda$,  i.e, $\Thetabold_{\ell,\lambda}(\matX) = (\matU,\matV)$.  We assume that the factorization rank and the regularizer parameter are set in a validation phase and remain fixed afterwards and we use $\Thetabold(\matX)$ and $\Thetabold_{\ell,\lambda}(\matX)$ interchangeably throughout the paper.

We use $R$ to denote the socially relevant objective function that we seek to optimize by adding antidote data. $R$ is a function of estimated ratings $\hat{\matX}$ and possibly (depending on the objective) other parameters such as original ratings, user labels, etc. For example, consider an objective that minimizes the difference of average estimation errors between two groups of users. In that case, $R$ is a function defined over $\matX$, $\hat{\matX}$, and another parameter that indicates the group membership of each user.  The specific objective
 functions we study in this paper are presented in Section~\ref{sec:Utilities}. Now we can formally state the optimal antidote data problem:
\begin{problem}[{Optimal Antidote Problem}]\label{antidote-problem}
Given a partially observed rating matrix $\matX \in \R^{n \times d}$, a budget $n' = \alpha n$, a factorization algorithm $\Thetabold_{\ell,\lambda}$, and an objective function $R$, find the antidote data $\tilde{\matX} \in \R^{n' \times d}$ such that $R$ is optimized when $\Thetabold_{\ell,\lambda}$ is applied jointly on $\matX$ and $\tilde{\matX}$.
\end{problem}
Note that we may want to either maximize or minimize $R$ depending on the objective.  Also, although in our notation $\tilde{\matX}$ corresponds to a set of artificial users, we can apply problem \ref{antidote-problem} to generate a set of artificial items by using the symmetry of the problem, i.e., by transposing $\matX$.

Although some objective functions have additional parameters such as the original observed ratings ($\matX$) or a list of group memberships (which we denote $K$), adding antidote data only affects the output of the factorization algorithm and hence the rating estimations $\hat{\matX}$. Therefore, we denote the general objective function by $R(\hat{\matX})$ instead of $R(\hat{\matX},\matX,K)$ for notational convenience. Assuming our goal is to minimize some objective function $R$, we can rewrite problem \ref{antidote-problem} as:
\begin{equation}
\label{eq:antidote-obj}
\underset{\tilde{\matX} \in \mathbb{M}}{\argmin} \quad R(\hat{\matX})
\end{equation}
where $\mathbb{M} \subset \R^{n' \times d}$ is the set of feasible antidote data matrices. 

Let $\Thetabold(\matX;\tilde{\matX})$ denote the factorization algorithm when applied jointly on the original and the antidote data. In this case, the output consists of the item latent vectors forming the columns of factor $\matV \in \R^{\ell \times d}$, and the user latent vectors which can be split into a matrix of original users latent vectors $\matU \in \R^{\ell \times n}$, and a matrix of antidote users latent vectors $\tilde{\matU} \in \R^{\ell \times n'}$; therefore, we have $\Thetabold(\matX;\tilde{\matX}) = (\matU,\tilde{\matU},\matV)$.

Furthermore, $\hat{\matX}$ is a function of original users latent vectors and item latent vectors\footnote{Note that here $\matU$ and $\matV$ are the users and items latent vectors after adding the antidote data to the system, which can be different from initial $\matU$ and $\matV$.}, i.e., $\hat{\matX}(\Thetabold(\matX;\tilde{\matX})) = \matU^\intercal \matV$. This allows us to write (\ref{eq:antidote-obj}) in the explicit form:
\begin{equation}
\label{eq:antidote-obj-explicit}
\underset{\tilde{\matX} \in \mathbb{M}}{\argmin} \quad R(\hat{\matX}(\Thetabold(\matX;\tilde{\matX})))
\end{equation}
In other words, we are looking for antidote data $\tilde{\matX}$ that modifies the outputs of $\Thetabold$ such that $\hat{\matX}$ is modified to optimize $R$. Figure \ref{fig:model} shows a schematic representation of the antidote data effect on matrix factorization models. In the next section, we introduce an iterative method to solve (\ref{eq:antidote-obj-explicit}).
\section{Computing Antidote Data}\label{sec:Method}
In this section we introduce the framework for generating antidote data. We apply a projected gradient descent/ascent algorithm (\textbf{\texttt{GD}}/\textbf{\texttt{GA}}) to optimize the antidote data with respect to a socially relevant objective function. In section \ref{sec:Method-recap} we review a gradient descent method, introduced in \cite{li2016data}, for optimizing data poisoning attacks on matrix factorization models, and which we adapt to optimize antidote data. Then, in section \ref{sec:Method-fast} we show how the characteristics of the antidote problem can be exploited for significant improvements in algorithmic efficiency.

\subsection{A Projected Gradient Descent Approach}\label{sec:Method-recap}
In this section we describe a projected gradient descent algorithm to solve the constrained optimization problem (\ref{eq:antidote-obj}). A parallel approach is taken in \cite{li2016data} for optimizing data poisoning attacks, which is itself an instance of the more general machine teaching problem introduced in \cite{mei2015using}. We note that the framework introduced in \cite{mei2015using} can be used to extend the applicability of antidote data approach beyond matrix factorization models.

The algorithm starts from an initial antidote data with size of a given budget. At each iteration, the factorization algorithm is applied jointly on the original data and the current antidote data to find updated factors $\matU$,$\tilde{\matU}$,$\matV$, and estimated ratings $\hat{\matX}$. Then the gradient of the antidote utility with respect to antidote data at the current point is computed and the algorithm chooses a step size and updates the antidote data. After each update, a projection function is applied to get a feasible solution. In this paper we only consider range constraints on the ratings, i.e., for each rating $\tilde{x}_{ij}$ we assume $\mathbb{M}_{min}<\tilde{x}_{ij}<\mathbb{M}_{max}$ where $\mathbb{M}_{min}$ and $\mathbb{M}_{max}$ indicate the minimum and maximum feasible rating in the system. Therefore the projection function simply truncates all the ratings in $\tilde{\matX}$ at $\mathbb{M}_{min}$ and $\mathbb{M}_{max}$.

Algorithm \ref{alg:compute-antidote-data} presents the details of our antidote data optimization method. If the goal is to maximize $R$, we can apply a gradient ascent algorithm by simply changing the sign of the gradient step in line 6. 
The learning algorithm $\Thetabold_{\ell,\lambda}$ is an input to Algorithm \ref{alg:compute-antidote-data}. This is a realistic assumption in a white-box scenario, i.e., a party with the full knowledge of the recommender system seeks to generate antidote data, which is an important case.  However, we emphasize that there are settings in which other parties with only partial knowledge of the system can successfully adopt the antidote data approach as well.  First of all, recent work  \cite{wang2018stealing} introduces a method for estimating the hyper-parameters of a learning algorithm.  Using that method we need not input $\Thetabold_{\ell,\lambda}$ to Algorithm \ref{alg:compute-antidote-data}, instead only providing the original factors $(\matU,\matV)$.  Moreover, in Section \ref{sec:Experiments} we introduce heuristic algorithms that require less information about the recommender system than does Algorithm \ref{alg:compute-antidote-data}. 

{\small
\SetKwInOut{init}{Initialization}
\begin{algorithm}
 \KwIn{Observed ratings $\matX \in \R^{n \times d}$, budget $n'$, factorization algorithm $\Thetabold_{\ell,\lambda}$, utility $R$, feasible set $\mathbb{M}$}
 \KwOut{Antidote data $\tilde{\matX}$}
 \init{initialize $\tilde{\matX}^{(0)} \in \R^{n' \times d}$, $t=0$}
 \BlankLine
 \While{convergence}{
     $\matU,\tilde{\matU},\matV = \Thetabold(\matX;\tilde{\matX}^{(t)})$\\
     $\hat{\matX} = \matU^\intercal \matV$\\
     Compute $\nabla_{\tilde{\matX}} R(\hat{\matX})$\\
     Find step size $\alpha$\\
     $\tilde{\matX}^{(t+1)} = \tilde{\matX}^{(t)} - 
              \alpha \nabla_{\tilde{\matX}} R(\hat{\matX})$\\
    $\tilde{\matX}^{(t+1)} = P_{\mathbb{M}}(\tilde{\matX}^{(t+1)})$\\
     $t \leftarrow t + 1$
 }
 \KwRet{$\tilde{\matX}^{(t)}$}
 \BlankLine
 \caption{Optimizing antidote data via projected gradient descent}\label{alg:compute-antidote-data}
\end{algorithm}
}
In order to compute $\nabla_{\tilde{\matX}} R(\hat{\matX})$ in line 4 of algorithm \ref{alg:compute-antidote-data}, we consider the explicit form of the objective function given in (\ref{eq:antidote-obj-explicit}). Applying the chain rule we get:
\begin{equation}
\begin{aligned}
\nabla_{\tilde{\matX}} R(\hat{\matX}) 
&= \nabla_{\Thetabold} R(\hat{\matX}) \ 
\nabla_{\tilde{\matX}} (\Thetabold(\matX;\tilde{\matX}))
\end{aligned}
\end{equation}

$\nabla_{\tilde{\matX}} (\Thetabold(\matX;\tilde{\matX}))$ is the Jacobian matrix that contains partial derivatives of factors $(\matU,\tilde{\matU},\matV)$ with respect to each element in $\tilde{\matX}$. These partial derivatives can be approximately computed by exploiting the KKT conditions of the factorization problem as explained in \cite{li2016data} \cite{mei2015using}. However, in section \ref{sec:Method-fast} we show cases where the full computation of such partial derivatives is not required and we explain how to derive the necessary elements.

By applying the chain rule one more time on $\nabla_{\Thetabold} R(\hat{\matX})$ we get:
\begin{equation}\label{eq:gradient}
\begin{aligned}
\nabla_{\tilde{\matX}} R(\hat{\matX}) 
&= \nabla_{\hat{\matX}} R(\hat{\matX}) \
\nabla_{\Thetabold} \hat{\matX}(\Thetabold) \
\nabla_{\tilde{\matX}} (\Thetabold(\matX;\tilde{\matX}))
\end{aligned}
\end{equation}
The first term in (\ref{eq:gradient}) is the gradient of the antidote utility with respect to the estimated ratings. In this paper we only consider differentiable utilities, as described in more detail in section \ref{sec:Utilities}.

The second term in (\ref{eq:gradient}) is the gradient of the estimated ratings with respect to factors $(\matU,\tilde{\matU},\matV)$. This term is straighforward to compute since the ratings are linear in each factor, i.e., $\hat{\matX} = \matU^\intercal \matV$.

In this paper we do not make assumptions (e.g. convexity) about the antidote utility other than being differentiable; the framework is a general method to improve a socially relevant metric rather than one that seeks the global optimum of function $R$. However, we note that introducing antidote objectives with certain provable properties, which can provide convergence guarantees or more efficient ways to find the step size in Algorithm \ref{alg:compute-antidote-data}, is a potential direction for future research.

\subsection{Efficient Computation of the Gradient Step}\label{sec:Method-fast}
In this section we show how to further simplify (\ref{eq:gradient}) to make the update step of Algorithm \ref{alg:compute-antidote-data} more efficient.

First, we write $\nabla_{\Thetabold} \hat{\matX}(\Thetabold)$ in terms of the block matrices that contain the partial derivatives of the estimated ratings in $\hat{\matX}$ with respect to each factor $\matU,\tilde{\matU},\matV$, i.e.%
\footnote{For matrices $\matA \in \R^{m\times n}$ and $\matB \in \R^{r\times s}$, we use $\frac{\partial \matA}{\partial \matB}$ to denote an $mn \times rs$ matrix that contains the partial derivatives $\frac{\partial a_{ij}}{\partial b_{k\ell}}$ for each $a_{ij}$ and $b_{k\ell}$.}%
, $\left[\frac{\partial \hat{\matX}}{\partial \matU}, \frac{\partial \hat{\matX}}{\partial \tilde{\matU}}, \frac{\partial \hat{\matX}}{\partial \matV}\right]$. Notice that $\hat{\matX} = \matU^\intercal \matV$ does not depend on $\tilde{\matU}$ and therefore $\frac{\partial \hat{\matX}}{\partial \tilde{\matU}}= \mathbf{0}$.

Furthermore, we write $\nabla_{\tilde{\matX}} (\Thetabold(\matX;\tilde{\matX}))$ in terms of the block matrices that contain the partial derivatives of each  factor $\matU,\tilde{\matU},\matV$ with respect to each element in $\tilde{\matX}$, i.e.,
$\left[
(\frac{\partial \matU}
{\partial \tilde{\matX}})^\intercal, 
(\frac{\partial \tilde{\matU}}
{\partial \tilde{\matX}})^\intercal,
(\frac{\partial \matV}
{\partial \tilde{\matX}})^\intercal
\right]^\intercal$. Assuming that an infinitesimal change in $\tilde{x}_{ij}$ only results in first order updates in vectors $\tilde{\vecu}_i$ and $\vecv_j$, we get $\frac{\partial \matU}{\partial \tilde{\matX}} = \mathbf{0}$.

Exploiting the fact that $\frac{\partial \matU}{\partial \tilde{\matX}} = \frac{\partial \hat{\matX}}{\partial \tilde{\matU}} = \mathbf{0}$, we can simplify (\ref{eq:gradient}) to:
\begin{equation}\label{eq:gradient-simple}
\begin{aligned}
\nabla_{\tilde{\matX}} R(\hat{\matX}) 
&= \nabla_{\hat{\matX}} R(\hat{\matX}) \ 
   \frac{\partial \hat{\matX}}{\partial \matV} \ 
   \frac{\partial \matV}{\partial \tilde{\matX}}  
\end{aligned}
\end{equation}
Now we derive $\frac{\partial R(\hat{\matX})}{\partial \tilde{x}_{ij}}$ for each element of the antidote data $\tilde{x}_{ij}$. Let $\vecv_1, \dots, \vecv_d$ be the item vectors forming the columns of $V$. Then starting from the last term in (\ref{eq:gradient-simple}) and assuming first order updates, we know that $\frac{\partial \vecv_k}{\partial \tilde{x}_{ij}}$ is non-zero only if $k=j$ and  can be approximately computed as\footnote{Details are provided in appendix A.1.}:
\begin{equation}
\frac{\partial \vecv_j}{\partial \tilde{x}_{ij}} =
(\sum_{i \in \Omega_j} \vecu_i \vecu^\intercal_i + \tilde{\matU} \tilde{\matU}^\intercal + 
\lambda \matI_{\ell})^{-1} \tilde{\vecu}_i
\end{equation}
On the other hand, $\frac{\partial \hat{x}_{lk}}{\partial \vecv_j} = \vecu_l^\intercal$ if $k=j$ and an $\ell$-dimensional zero vector otherwise. Therefore, we need to compute $\frac{\partial R(\hat{\matX})}{\partial \hat{x}_{lk}}$ only for $k=j$ and we have:
\begin{equation}\label{eq:gradient-efficient}
\frac{\partial R(\hat{\matX})}{\partial \tilde{x}_{ij}} = \left(\sum_{l=1}^n  \frac{\partial R(\hat{\matX})}{\partial \hat{x}_{lj}} 
\frac{\partial \hat{x}_{lj}}{\partial \vecv_j}\right)
\frac{\partial \vecv_j}{\partial \tilde{x}_{ij}}
\end{equation}
Let $\matG$ be a matrix formed by reshaping  $\nabla_{\hat{\matX}} R(\hat{\matX})$ into an $n \times d$ matrix such that $g_{ij} = \frac{\partial R(\hat{\matX})}{\partial \hat{x}_{ij}}$. Then we can write (\ref{eq:gradient-efficient}) as:
\begin{equation}\label{eq:gradient-final}
\frac{\partial R(\hat{\matX})}{\partial \tilde{x}_{ij}} = {\vecg_j}^\intercal \matU^\intercal \matS_j^{-1} \tilde{\vecu}_i
\end{equation}
where $\matS_j = \sum_{i \in \Omega_j} \vecu_i \vecu^\intercal_i + \tilde{\matU} \tilde{\matU}^\intercal + 
\lambda \matI_{\ell}$.

By using (\ref{eq:gradient-final}) instead of the general formula in (\ref{eq:gradient}) we can significantly reduce the number of computations required for finding the gradient of the utility function with respect to the antidote data. Furthermore, the term ${\vecg_j}^\intercal \matU^\intercal \matS_j^{-1}$ appears in all the partial derivatives that correspond to elements in column $j$ of $\tilde{\matX}$ and can be precomputed in each iteration of the algorithm and reused for computing partial derivatives with respect to different antidote users.

\section{Social Objective Functions}\label{sec:Utilities}
The previous section developed a general framework for improving various
properties of recommender systems;  in this section we show how to apply
that framework specifically to issues of polarization and fairness.


As described in Section~\ref{sec:relwork}, polarization is the degree to which opinions, views, and sentiments
  diverge within a population.   Recommender systems can capture this
  effect through the ratings that they present for items.  To formalize
  this notion, we define polarization in terms of the variability of
  predicted ratings when compared across users.   In fact, we note that
  both very high variability, and very low variability of ratings may be
  undesirable.   In the case of high variability, users have strongly
  divergent opinions, leading to conflict.   Recent analyses of the YouTube recommendation system have suggested that it can enhance this effect \cite{Nicas:WSJ18,ocallaghan15:down_white_rabbit_hole}.  On the other hand, 
the convergence of user preferences, i.e., very low variability of ratings given to each item across users, corresponds to increased homogeneity, an undesirable phenomenon that may occur as users interact with a recommender system~\cite{chaney2017algorithmic}. As a result, in what follows we consider using antidote data in both ways: to either increase or decrease polarization.

As also described in Section~\ref{sec:relwork}, unfairness is a topic of growing interest in machine learning. Following the discussion in that section, we consider a recommender system fair if it provides equal quality of service (i.e., prediction accuracy) to all users or all groups of users \cite{zafar2017fairness}.

Next we formally define the metrics that specify the objective
 functions associated with each of the above objectives. Since the gradient of each objective function is used in the optimization algorithm, for reproducibility we provide the details about derivation of the gradients in appendix A.2.

\subsection{Polarization}\label{sec:Utilities-polarization}
To capture polarization, we seek to measure the extent to which the user ratings \emph{disagree}.
Thus, to measure user polarization we consider the estimated ratings $\hat{\matX}$, and we define the polarization metric as the normalized sum of pairwise euclidean distances between estimated user ratings, i.e., between rows of $\hat{\matX}$. In particular:
\begin{equation}\label{polarization-obj}
R_{pol}(\hat{\matX}) = \frac{1}{n^2 d} \sum_{k=1}^n \sum_{l>k}
||\hat{\vecx}^k - \hat{\vecx}^l||^2
\end{equation}

The normalization term $\frac{1}{n^2 d}$ in (\ref{polarization-obj}) makes the polarization metric identical to the following definition:
\footnote{We can derive it by rewriting (\ref{polarization-obj}) as $\displaystyle R_{pol}(\hat{\matX}) = \frac{1}{d} \sum_{j=1}^d \frac{1}{n^2} \sum_{k=1}^n \sum_{l>k}
 (\hat{x}_{kj} - \hat{x}_{lj})^2$.}
\begin{equation}\label{polarization-obj-2}
R_{pol}(\hat{\matX}) = \frac{1}{d} \sum_{j=1}^d \sigma_j^2
\end{equation}
where $\sigma_j^2$ is the variance of estimated user ratings for item $j$. Thus this polarization metric can be interpreted either as the average of the variances of estimated ratings in each item, or equivalently as the average user disagreement over all items.

\subsection{Fairness}\label{sec:Utilities-fairness}
\spara{Individual fairness.} 
For each user $i$, we define $\ell_i$, the loss of user $i$, as the mean squared estimation error over known ratings of user $i$:
\begin{equation}
\ell_i = \frac{||P_{\Omega^i} (\hat{\vecx}^i - \vecx^i)||_2^2}{|\Omega^i|}
\end{equation}
Then we define the individual unfairness as the variance of the user losses:\footnote{Note that for a set of equally likely values $x_1,\dots,x_n$ the variance can be expressed without referring to the mean as: $\frac{1}{n^2}\displaystyle\sum_i \displaystyle\sum_{j>i}(x_i - x_j)^2.$}
\begin{equation}\label{individual-fairnss-obj}
R_{indv}(\matX,\hat{\matX}) = \frac{1}{n^2} \sum_{k=1}^n \sum_{l>k} (\ell_k - \ell_l)^2 
\end{equation}
To improve individual fairness, we seek to minimize $R_{indv}$.

\spara{Group fairness.}
Let $I$ be the set of all users/items and $G=\{G_1\, \dots, G_g\}$ be a partition of users/items into $g$ groups, i.e., $I= \bigcup_{i \in \{1, \dots, g\}} G_i$. We define the loss of group $i$ as the mean squared estimation error over all known ratings in group $i$:

\begin{equation}
L_i = \frac{||P_{\Omega_{G_i}} (\hat{\matX} - \matX)||_2^2}{|\Omega_{G_i}|}
\end{equation}
For a given partition $G$, we define the group unfairness as the variance of all group losses:
\begin{equation}\label{group-fairnss-obj}
R_{grp}(\matX,\hat{\matX},G) = \frac{1}{g^2} \sum_{k=1}^g \sum_{l>k} (L_k - L_l)^2
\end{equation}
Again, to improve group fairness, we seek to minimize $R_{grp}$. 

\subsection{Accuracy vs. Social Welfare }
Adding antidote data to the system to improve a social utility will also have an effect on the overall prediction accuracy. Previous works have considered social objectives as regularizers or constraints added to the recommender model (eg, \cite{burke2018balanced, zafar2017fairness2, kamishima2011fairness}), implying a trade-off between the prediction accuracy and a social objective.

However, in the case of the metrics we define here, the relationship is not as simple. Considering polarization, we find that in general, increasing or decreasing polarization will tend to decrease system accuracy. In either case we find that system accuracy only declines slightly in our experiments; we report on the specific values in Section~\ref{sec:Experiments}. Considering either individual or group unfairness, the situation is more subtle. Note that our unfairness metrics will be exactly zero for a system with zero error (perfect accuracy).  As a result, it is possible that as the system decreases unfairness, overall accuracy may either increase or decrease. We illustrate these effects in our experiments in Section~\ref{sec:Experiments}.


\section{Effectiveness}\label{sec:Experiments}
\begin{figure}[t]
    \centering
    \begin{subfigure}[]{0.236\textwidth}
        \includegraphics[width=\textwidth]{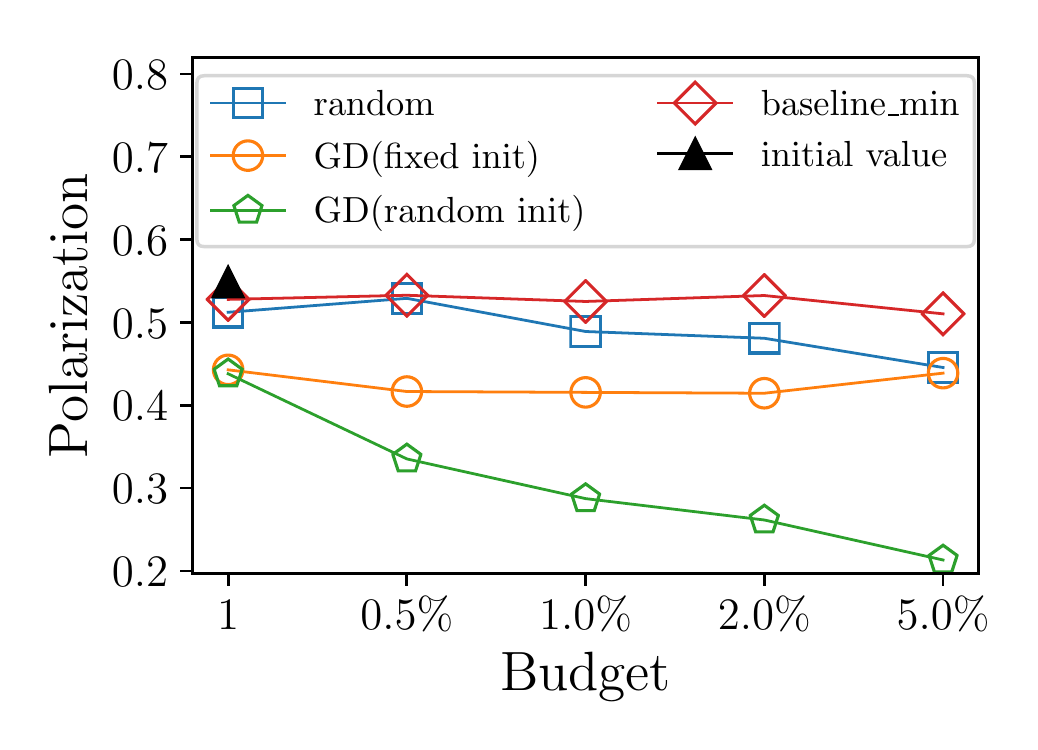}
        \vspace*{-\baselineskip}
        \caption{\small Minimizing polarization}
        \label{fig:polmin}
    \end{subfigure}
    \begin{subfigure}[]{0.236\textwidth}
        \includegraphics[width=\textwidth]{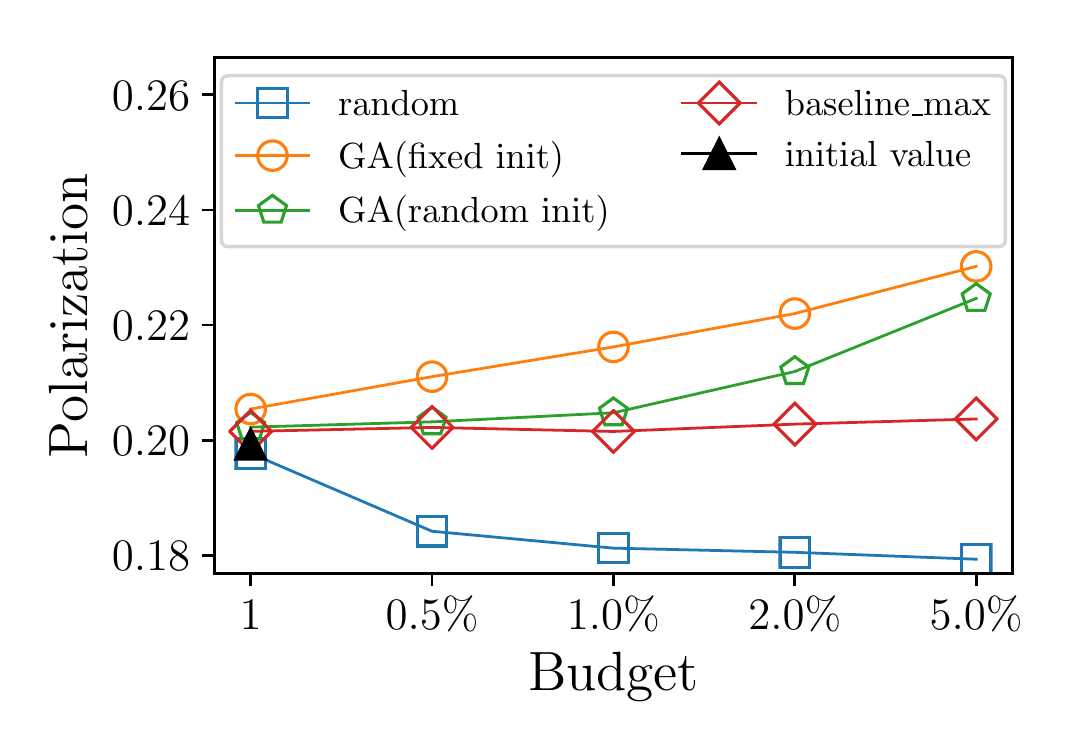}
        \vspace*{-\baselineskip}
        \caption{\small Maximizing polarization}
        \label{fig:polmax}
    \end{subfigure}
    \vspace*{-\baselineskip}
    \caption{\small Modifying user polarization.}\label{fig:pol}
\end{figure}

In this section we use the tools developed in previous sections to study the effectiveness of antidote data in varying the polarization and reducing the unfairness of a matrix-factorization based recommender system.

We consider a recommender system that estimates unknown ratings by solving the regularized matrix factorization problem as defined by (\ref{eq:MFobj}).  We implemented an alternating least squares algorithm \cite{hastie2015matrix,hardt2014understanding} to find the factors. We use the MovieLens 1M dataset which contains around 1 million ratings of $\sim$4000 movies made by $\sim$6000 users, with ratings on a 5-point scale \cite{harper2016movielens}. We choose the 1000 most frequently rated movies, and use different subsets of users in different experiments as described below.

For each dataset we perform a validation process to choose the hyper-parameters $(\ell,\lambda)$ so as to obtain realistic settings. The hyper-parameters are selected based on the average root-mean-square error (RMSE) of the factorization in multiple random splits of observed ratings into training and validation sets. We assume that the hyper-parameters are fixed during the antidote data generation process since the antidote data is generated for a fixed recommender system.

First we show the effectiveness of antidote data in modifying the user polarization as defined in section \ref{sec:Utilities-polarization}. In section \ref{sec:heuristics} we describe different heuristics that can significantly speed up the construction of antidote data. Finally, section \ref{sec:fairness} demonstrates the effectiveness of applying antidote data for improving fairness.

\subsection{Polarization}\label{sec:pol}
To explore modifying user polarization, we choose a random subset of 1000 users yielding a matrix in which 11\% of the elements are known.
As previously mentioned, it may be of interest to either increase or decrease the polarization metric in different scenarios.  We present an example for each case.  We do so by taking advantage of the fact that different hyperparameter combinations can yield models that are very close in overall accuracy but that differ significantly with respect to initial user polarization in the system.

In particular, we observe that the average validation RMSE over ten random splits of observed ratings into training and validation sets for $(rank=8,\lambda=10)$ is $\mathbf{0.87}$ and for $(rank=4,\lambda=0.1)$ is $\mathbf{0.90}$. However, the polarization ($R_{pol}$) of the estimated rating matrix for $(rank=8,\lambda=10)$ is $0.2$ whereas for $(rank=4,\lambda=0.1)$ the polarization goes up to $0.55$. We use the former setting as an example where the goal is to increase the polarization metric (to avoid homogeneity), and the latter setting as an example of a polarized system where the goal is to reduce polarization.

For each of the maximization and minimization objectives, we compare the performance of the antidote data generation framework with a baseline algorithm.  When seeking to minimize polarization, we use \textbf{\texttt{baseline\textunderscore min}}. This algorithm tries to reduce the variance of estimated ratings in each item by setting the ratings given to the corresponding item in the antidote data to the average of known ratings for that item in the original data.  When seeking to maximize polization, we use 
\textbf{\texttt{baseline\textunderscore max.}} This algorithm generates antidote data by setting half of the user ratings in each item to the maximum feasible rating value and the other half of the ratings to the minimum feasible rating value. 

Furthermore, we consider two different initializations for the optimization process: in the case of \textbf{\texttt{GD(fixed init)}}, all the ratings in the initial antidote data are set to the same value. In the case of \textbf{\texttt{GD(random init)}}, we run the optimization multiple times starting from random initializations and return the best solution. 
\begin{figure}[t]
    \centering
    \begin{subfigure}[]{0.236\textwidth}
        \includegraphics[width=\textwidth]{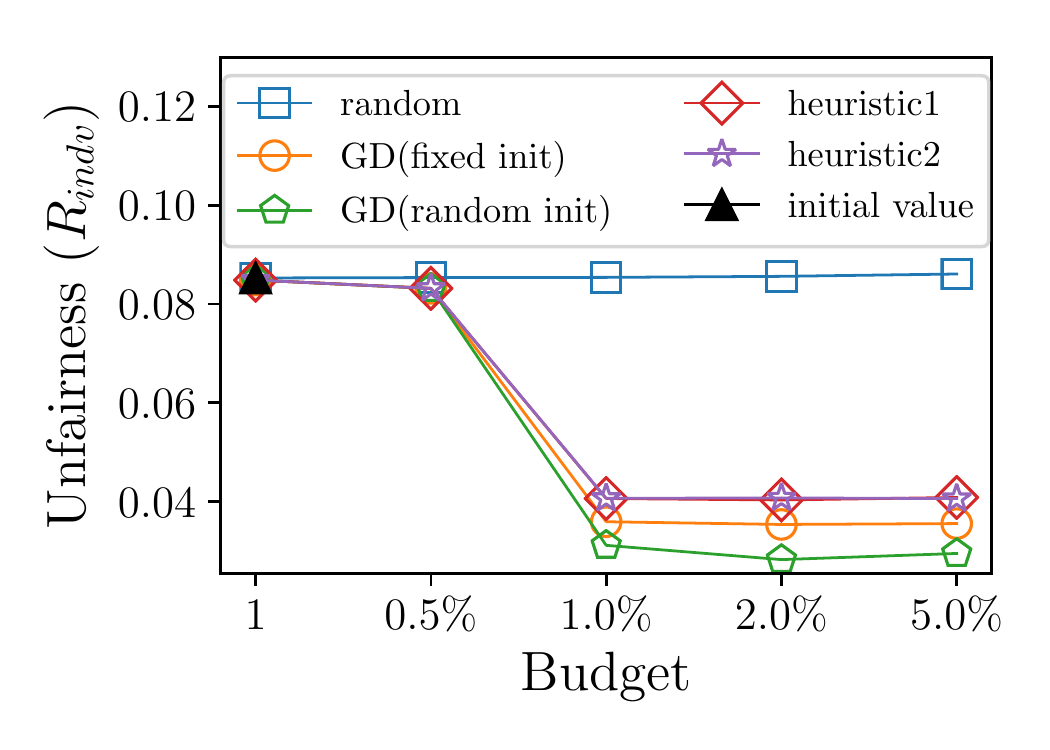}
        \vspace*{-\baselineskip}
        \caption{\small Individual fairness}
        \label{fig:individual-fairness}
    \end{subfigure}
    \begin{subfigure}[]{0.236\textwidth}
        \includegraphics[width=\textwidth]{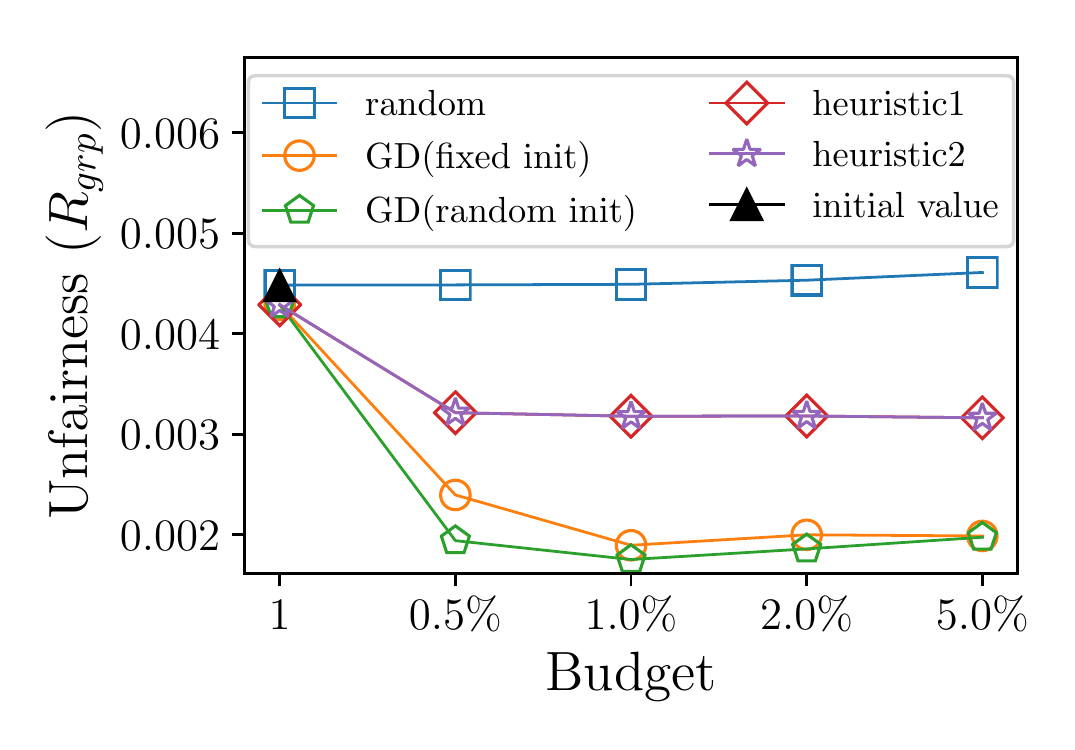}
		\vspace*{-\baselineskip}        
        \caption{\small Group fairness}
        \label{fig:group-fairness}
    \end{subfigure}
    \vspace*{-\baselineskip}
    \caption{\small Improving fairness.}\label{fig:fairness}
\end{figure}
Figure \ref{fig:pol} compares the effects of adding antidote data constructed by different methods on polarization. After each injection of the antidote data, the new polarization is computed using the original data only, i.e., we ignore the injected data in evaluating polarization. We present our results for different budgets varying from a single antidote user to 5\% of the number of original users. We also show the effect of ratings that are randomly generated over the feasible range, when used as antidote data.

\begin{figure*}[h]
    \centering
    \begin{subfigure}[]{0.26\textwidth}
        \includegraphics[width=\textwidth]{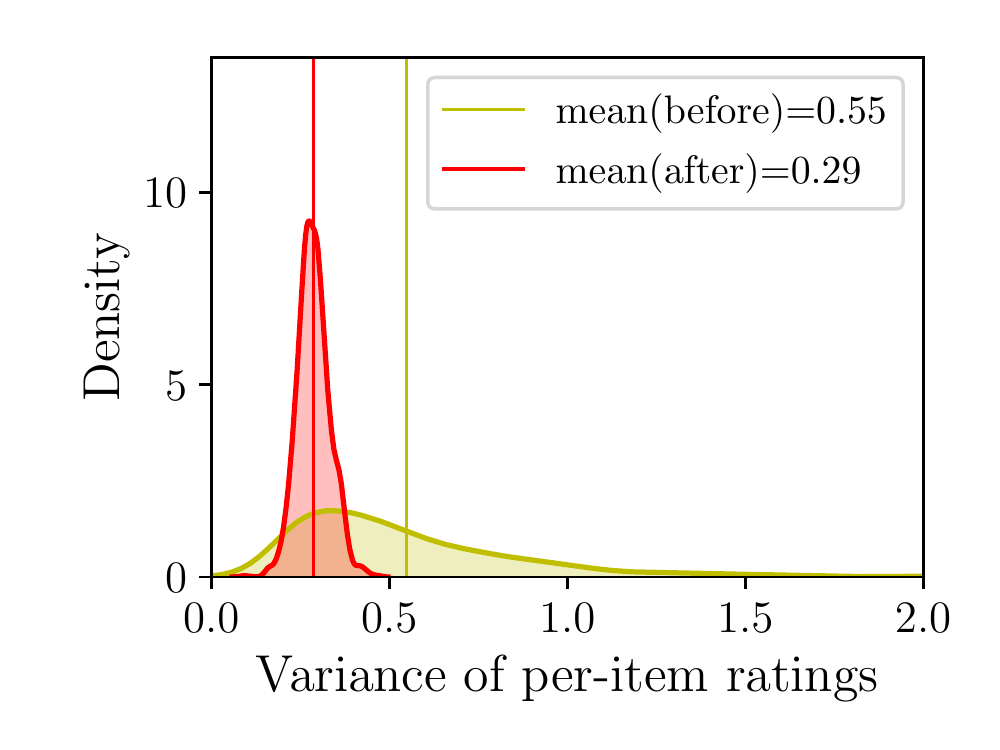}
        \vspace*{-\baselineskip}
        \caption{\small Effect on per-item polarization}
        \label{fig:polmin-example-a}
    \end{subfigure}
    \quad
    \begin{subfigure}[]{0.26\textwidth}
        \includegraphics[width=\textwidth]{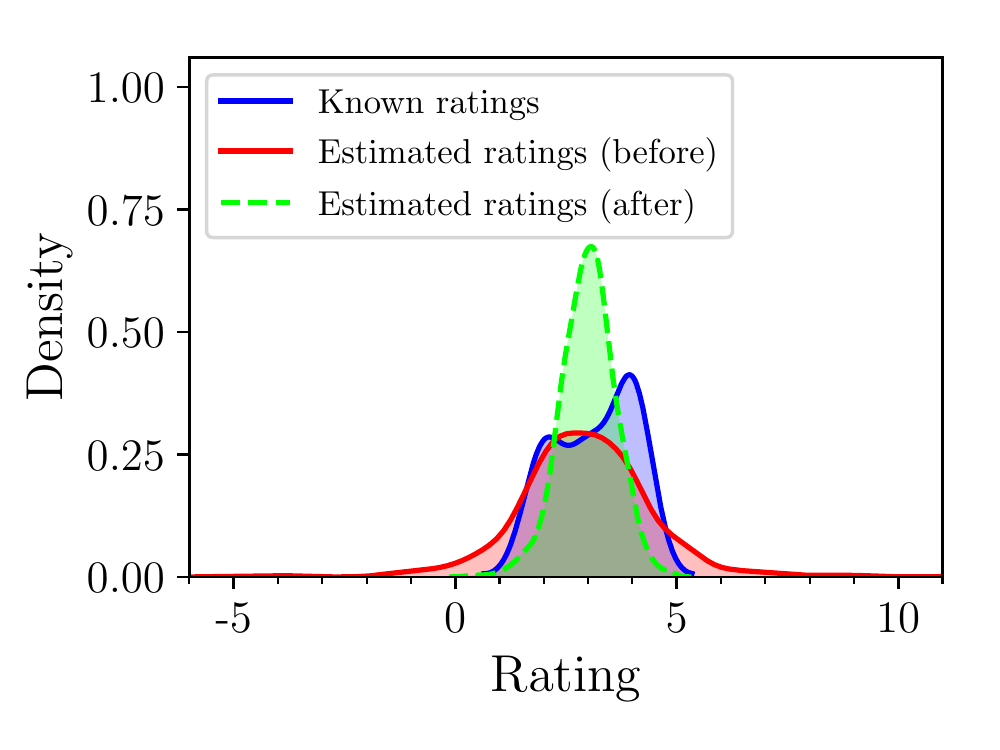}
        \vspace*{-\baselineskip}
        \caption{\small Effect on rating estimations for \emph{Patch Adams (1998)}}
        \label{fig:polmin-example-b}
    \end{subfigure}
    \quad
    \begin{subfigure}[]{0.26\textwidth}
        \includegraphics[width=\textwidth]{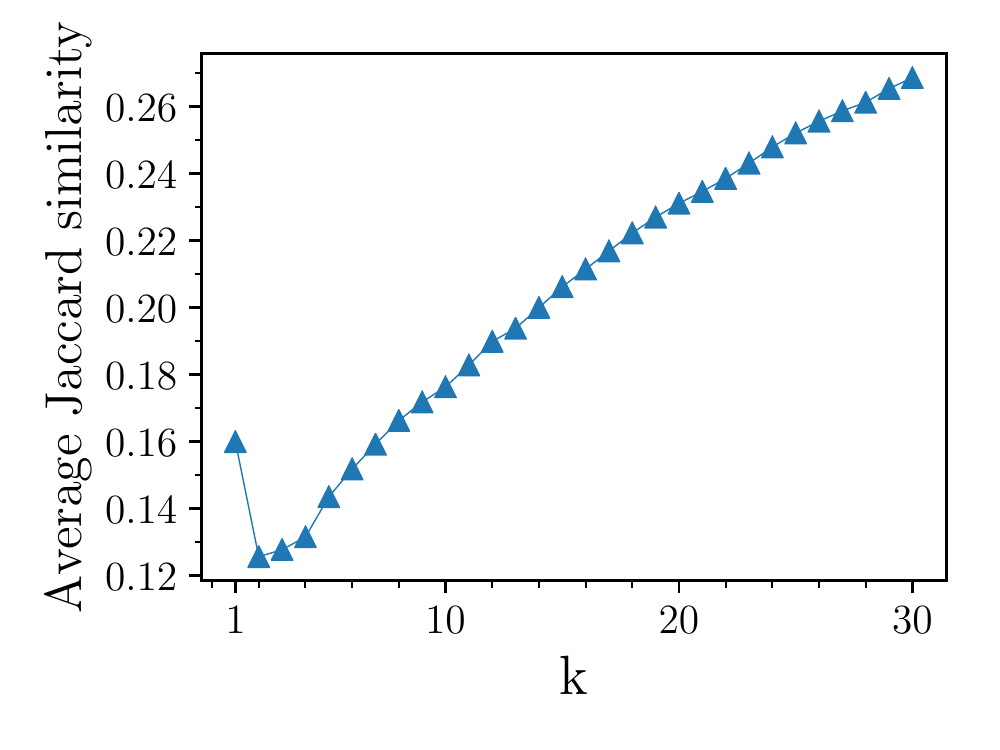}
        \vspace*{-\baselineskip}
        \caption{\small Effect on top-k recommended items}
        \label{fig:polmin-example-c}
    \end{subfigure}
    \vspace*{-\baselineskip}
    \caption{\small Minimizing polarization with $1\%$ budget.}\label{fig:polmin-example} 
\end{figure*}

Our results show that the antidote data generation framework can successfully either minimize or maximize polarization.  Antidote data generated by our method are considerably more effective than the baseline algorithms as well as random data. We observe that a 2\% budget is enough to reduce the initial polarization in a polarized setting by 50\% and increase the polarization in a less polarized setting by 10\%. Furthermore, we observe that random initialization is more effective for minimizing polarization whereas initializing all the antidote ratings from the same value is more effective for maximizing polarization.

To better understand the effect of antidote data on user polarization, in Figure \ref{fig:polmin-example} we demonstrate the effect of antidote data with a 1\% budget for the minimization case. Note that the effect on estimation error of adding antidote data is negligible: RMSE of rating estimations for known elements changes from $\mathbf{0.80}$ to $\mathbf{0.83}$.  In other words, antidote data modifies the prediction model such that its predictions still approximately agree with the known ratings but the polarization of the new estimated rating matrix is significantly different.

Figure \ref{fig:polmin-example-a} shows the distributions of per-item polarization  ($\sigma_j^2$ in (\ref{polarization-obj-2})) along with $R_{pol}$ (the distributional mean) before and after antidote data injection.   The figure shows that without antidote data, a small set of items make large contributions to overall polarization -- they have quite high variance in ratings, shown by the long distributional tail.  The addition of antidote data dramatically reduces this effect, and also significantly lowers $R_{pol}$ from 0.55 to 0.29.

Figure \ref{fig:polmin-example-b} shows the effect of adding antidote data on the estimated ratings of \emph{Patch Adams (1998)}, one of the movies for which the variance of estimated ratings is large before adding antidote data. We observe that the  distribution of known ratings for this movie indicates a polarized case with two peaks at 2 and 4. The initial rating estimations in this case lie in an interval that is much larger than the range of observed ratings. Adding antidote data modifies the extreme rating estimations; resulting in a unimodal distribution over the range of original ratings.

While the goal of adding antidote data is to modify the system's predicted ratings, an important use case for such a system is to output the top rated items as the system's recomendations.  Hence, it is important to ask how modifying predicted ratings will change the ranking of unrated items, i.e., the output of a top-$k$ recommender system. Therefore, we consider the top-$k$ recommended items on a per-user basis and measure the degree of change in the recommendations before and after adding antidote data. We use the Jaccard similarity of the sets of recommended items to measure this change.

Figure \ref{fig:polmin-example-c} shows the average of Jaccard similarities across all users.   
Our results show that the antidote data significantly changes the output of a top-$k$ recommender system.   For example, adding 1\% additional antidote data changes the top-recommended item for 84\% of all users.   We observe that, in general, as the number of considered top items grows, the effect lessens (Jaccard similarity grows). However, the changes in the set of recommended items are still significant up to $k=30$.

\subsection{Heuristic Algorithms}\label{sec:heuristics}
\begin{figure}[t]
    \centering
    \begin{subfigure}[]{0.236\textwidth}
        \includegraphics[width=\textwidth]{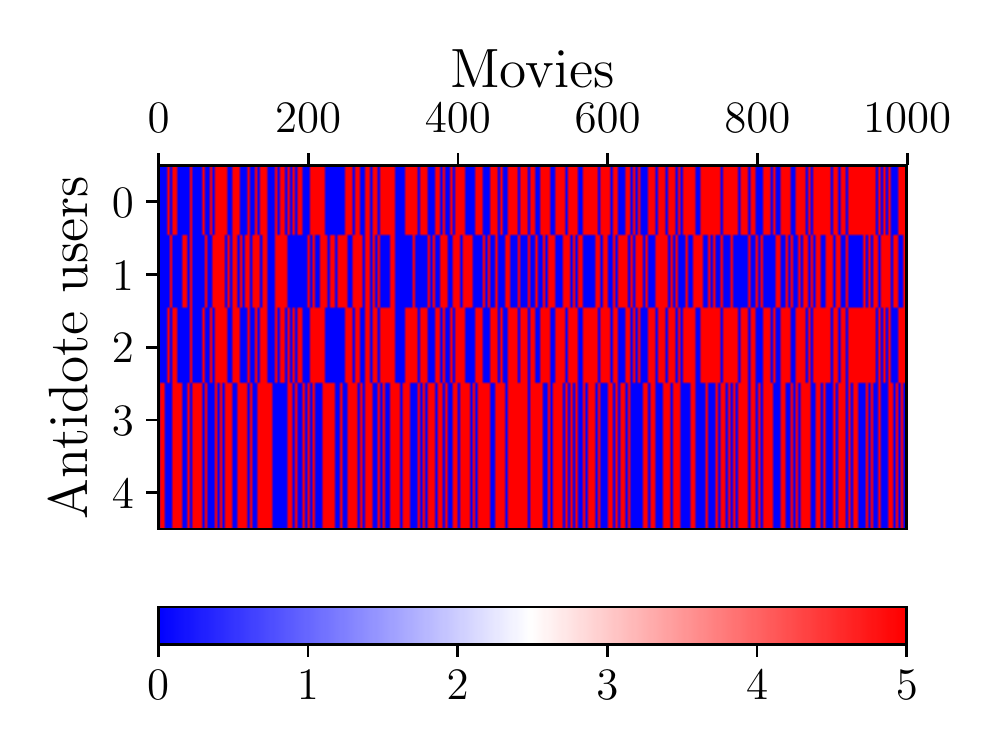}
        \vspace*{-\baselineskip}
        \caption{\small Minimizing polarization}
        \label{fig:pol-optimal}
    \end{subfigure}
    \begin{subfigure}[]{0.236\textwidth}
        \includegraphics[width=\textwidth]{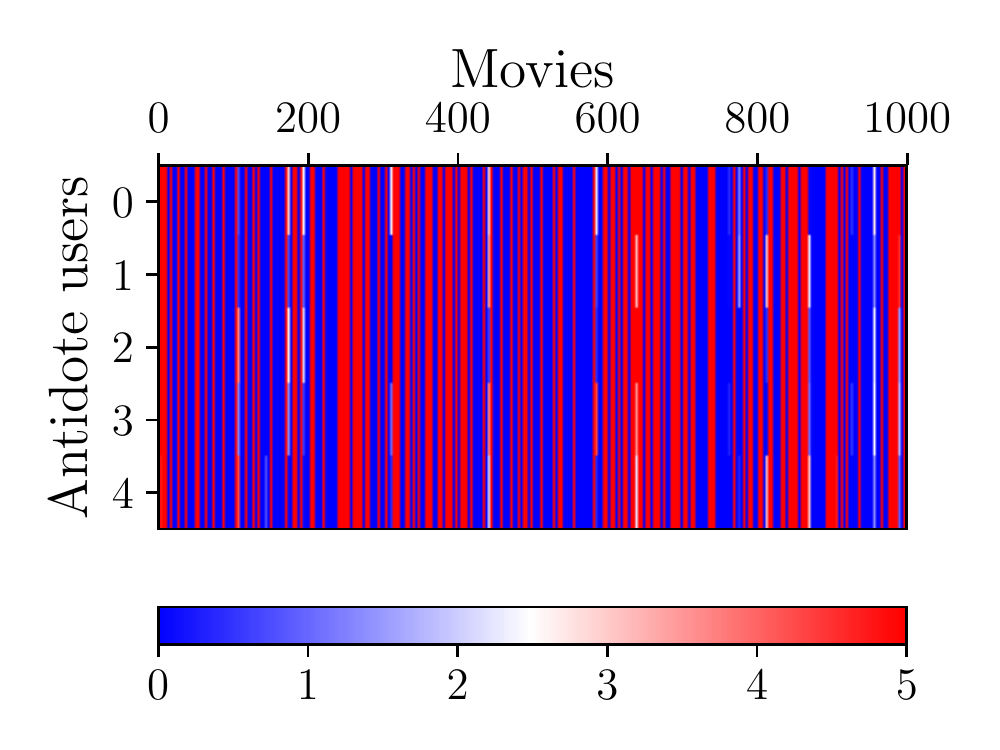}
		\vspace*{-\baselineskip}        
        \caption{\small Individual fairness}
        \label{fig:individual-optimal}
    \end{subfigure}
    \vspace*{-\baselineskip}
    \caption{\small Optimal antidote data with $0.5\%$ budget.}
    \label{fig:optimal}
\end{figure}

In this section we introduce heuristic algorithms that dramatically reduce the computational cost of antidote data generation. Notice that the computational cost of Algorithm 1 is dominated by performing the matrix factorization algorithm (evaluating $\Thetabold$) in each pass through the gradient descent loop. The heuristics are designed based on various approximations that can be made in different steps of Algorithm 1 to minimize the number of times $\Thetabold$ is evaluated. The approximations are motivated by certain patterns observed in the antidote data generated by Algorithm 1.

Figure~\ref{fig:optimal} shows the antidote data generated by \textbf{\texttt{GD(random init)}} for minimizing polarization (Fig.~\ref{fig:pol-optimal}) and minimizing individual unfairness (Fig.~\ref{fig:individual-optimal}). We observe that: (i) most of the ratings in the resulting antidote data are equal to one of the boundary values in the feasible set, $\mathbb{M}_{min}$ or $\mathbb{M}_{max}$ (0 or 5 in our experiments), and (ii) in the fairness case, most of the users (rows) in the antidote data converge to a nearly-identical pattern of ratings over items, even if they are initialized with different random values.

Based on the above observations, in our evaluations we consider two heuristics for generating antidote data for fairness.	The first (\textbf{\texttt{heuristic1}}) offers considerable computational savings, and the second (\textbf{\texttt{heuristic2}}) offers even more savings, while additionally removing the need for access to the factorization algorithm or its hyper-parameters ($\ell$,$\lambda$)\footnote{Pseudocodes are provided in appendix B.}.

\textbf{\texttt{heuristic1}} reduces the number evaluations of $\Thetabold$ to a single call by combining observations (i) and (ii).   It works by considering the addition of only a single row of antidote data, and computes gradients for that row.   Rather than performing gradient descent over a series of small steps, it then simply sets each value in the antidote data row to either $\mathbb{M}_{min}$ or $\mathbb{M}_{max}$ depending on the sign of the gradient.  It then replicates the resulting row as many times as dictated by the antidote data budget.

In the case of \textbf{\texttt{heuristic2}}, in addition to using the above observations, we approximate the direction of the gradient without the need to perform matrix factorization, given access to factors $\matU$ and $\matV$.    In this case, access to the factorization algorithm or its hyper-parameters is not required. Notice that (\ref{eq:gradient-final}) can be rewritten as $\frac{\partial R(\hat{\matX})}{\partial \tilde{x}_{1j}} = \veca_j^\intercal \vecb_j$ where $\veca_j=\left[\vecg_j^\intercal \matU^\intercal\right]^\intercal$ and $\vecb_j=\matS_j^{-1} \tilde{\vecu}_1$.    For sufficient level of regularization $\lambda$, we can approximate $\veca_j^\intercal \vecb_j \approx c \veca_j^\intercal \mathbf{1}_\ell$ where $c$ is a constant and $\mathbf{1}_\ell$ is an $\ell$-dimensional vector of $1$'s.   This leads to a modification of \textbf{\texttt{heuristic1}} in which all the values in column $j$ of the antidote data are set to 
$\mathbb{M}_{min}$ or $\mathbb{M}_{max}$ depending on the sign of $\vecg_j^\intercal \matU^\intercal \mathbf{1}_\ell$.

\subsection{Fairness}\label{sec:fairness}
\begin{table}[t]
  \caption{\small Effect of antidote data on individual unfairness $R_{indv}$ in the held-out ratings. $R_{indv}$ before antidote is 0.1087.}
  \label{tab:individual}
  \small
  \begin{tabular}{*{7}{l}}
    \toprule \toprule
     & \multicolumn{5}{c}{Budget}\\
    \cline{2-6}
    Algorithm & 1 & 0.5\% & 1\% & 2\% & 5\%\\
    \midrule
    GD(random init) & 0.1086 & 0.1084 & 0.1054 & 0.1157 & 0.1086 \\ 
	GD(fixed init) & 0.1086 & \textbf{0.1083} & 0.0929 & 0.0985 & 0.0968 \\ 
	heuristic1 & 0.1086 & 0.1084 & \textbf{0.0816} & \hl{\textbf{0.0800}} & 0.0830 \\ 
	heuristic2 & 0.1086 & 0.1084 & 0.0817 & 0.0818 & \textbf{0.0811} \\ 
  \bottomrule
\end{tabular}
\end{table}

In this section we show how antidote data as generated by our various algorithms improves fairness. We again use the MovieLens dataset; to study group fairness, we group movies by genre as specified in the dataset. In contrast to the case for polarization, the fairness objective is a function of both the known and predicted user ratings. Hence we choose the 1000 most active users and the 1000 most frequently rated movies. This gives us a rating matrix in which $\sim$36\% of the elements are known. For this dataset we run the matrix factorization algorithm with hyper-parameters $(rank=8,\lambda=1)$.

To verify that adding antidote data improves the fairness of unseen ratings, we hold out 20\% of the known ratings per user as a test set. We use the remaining data (training set) to generate antidote data; we then measure the effectiveness of the resulting antidote data in both training and test sets.

We start by assessing the effect of antidote data on fairness in the training data. We show the impact of antidote data on individual unfairness ($R_{indv}$) in Figure~\ref{fig:individual-fairness} and on group unfairness ($R_{grp}$) in Figure~\ref{fig:group-fairness}.  The figures compare the effect of antidote data as generated by four different algorithms: Algorithm~\ref{alg:compute-antidote-data} with two different initializations as described in Section~\ref{sec:pol}, and the two heuristic algorithms introduced in Section~\ref{sec:heuristics}.

The results show that all algorithms improve fairness considerably. In fact most of the benefits of antidote data can be obtained by only adding 1\% additional users in the individual fairness case and 0.5\% additional users in the group fairness case.    The figures also show that the much simpler heuristics, in which all rows of the antidote data are identical, are effective:  for individual fairness, they provide almost all the benefits of Algorithm~\ref{alg:compute-antidote-data} while for group fairness they provide around half of the benefits of Algorithm~\ref{alg:compute-antidote-data}.

\begin{table}[t]
  \caption{\small Effect of antidote data on group unfairness $R_{grp}$ in the held-out ratings. $R_{grp}$ before antidote is 0.0088.}
  \label{tab:group}
  \small
  \begin{tabular}{*{7}{l}}
    \toprule \toprule
     & \multicolumn{5}{c}{Budget}\\
    \cline{2-6}
    Algorithm & 1 & 0.5\% & 1\% & 2\% & 5\%\\
    \midrule
    GD(random init) & 0.0087 & \hl{\textbf{0.0035}} & 0.0041 & 0.0042 & 0.0045 \\ 
	GD(fixed init) & 0.0087 & 0.0042 & \textbf{0.0040} & \textbf{0.0040} & \textbf{0.0044} \\ 
	heuristic1 & 0.0087 & 0.0055 & 0.0056 & 0.0057 & 0.0058 \\ 
	heuristic2 & 0.0087 & 0.0055 & 0.0056 & 0.0057 & 0.0058 \\  
  \bottomrule
\end{tabular}
\end{table} 

Tables \ref{tab:individual} and \ref{tab:group} show the resulting values of the individual and group unfairness metrics in the \emph{test set} after antidote data addition for different budgets and different algorithms. We observe that the antidote data generated to reduce unfairness in the training data is also effective for reducing unfairness on the the held-out test data. The optimal value (minimum unfairness) in each table is highlighted.	We observe that even in the test set, a 2\% budget using \textbf{\texttt{heuristic1}} can reduce individual unfairness by over 25\% (from 0.1087 to 0.0800), and group unfairness can be lowered by more than 50\% (from 0.0088 to 0.0035) using \textbf{\texttt{GD(random init)}} and a 0.5\% budget.

\begin{figure}[h]
    \centering
    \begin{subfigure}[]{0.23\textwidth}
        \includegraphics[width=\textwidth]{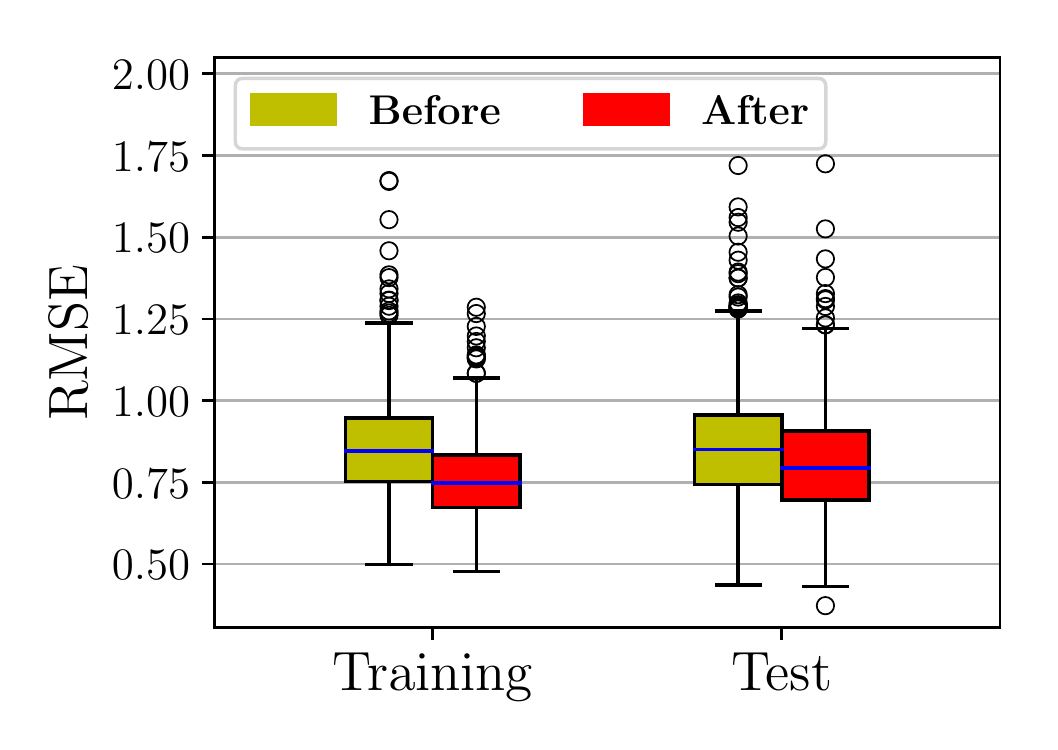}
        \vspace*{-\baselineskip}
        \caption{\small Individual fairness}
        \label{fig:indv-example}
    \end{subfigure}
    \begin{subfigure}[]{0.24\textwidth}
        \includegraphics[width=\textwidth]{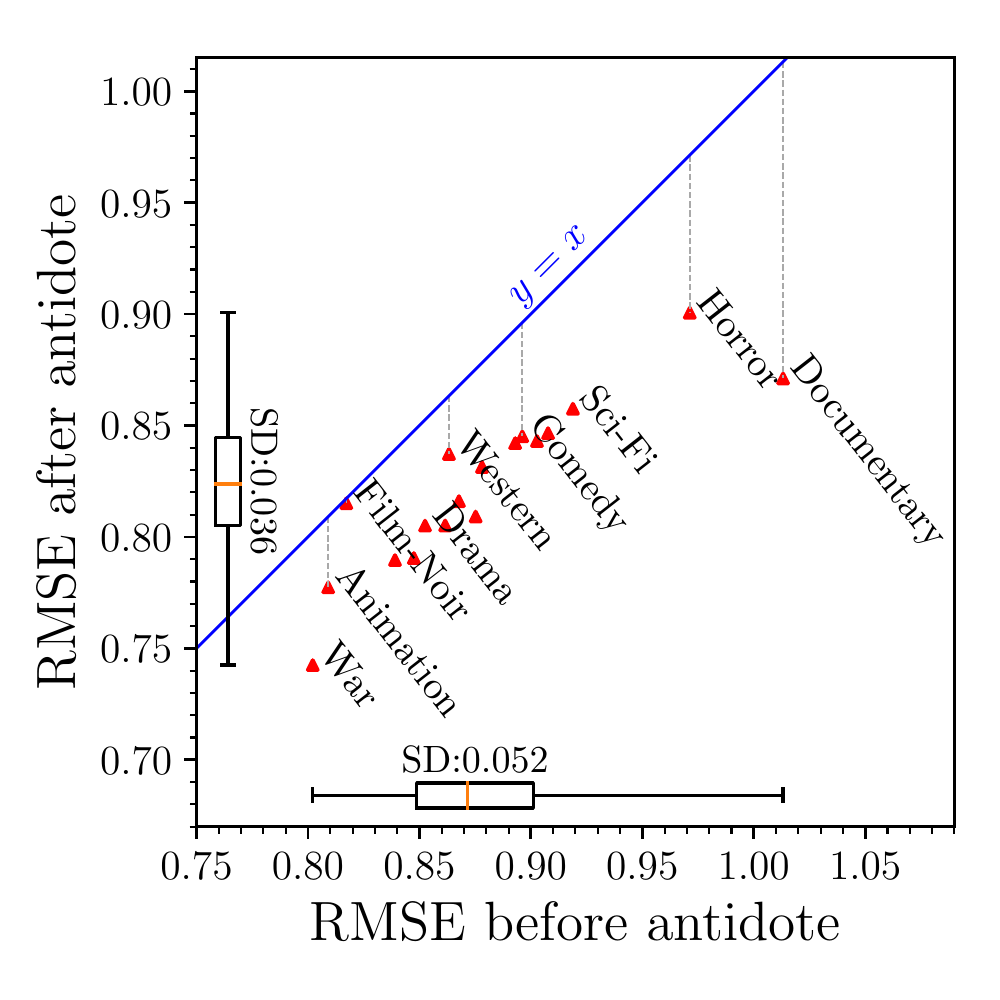}
        \vspace*{-\baselineskip}
        \caption{\small Group fairness}
        \label{fig:group-example}
    \end{subfigure}
    \vspace*{-\baselineskip}
    \caption{\small Antidote data effect on fairness.}
    \label{fig:fairness-example}
\end{figure}

Figure~\ref{fig:fairness-example} provides more insight into how adding antidote data reduces individual and group unfairness.
In each case, we consider the setting that reaches the minimum unfairness in the test set as presented in tables \ref{tab:individual} and \ref{tab:group}.

Figure~\ref{fig:indv-example} shows the effect of optimal antidote data on per-user RMSEs. The figure demonstrates a number of points.   First, adding antidote data results in a model with less variation in per-user RMSE of rating estimations in both training and test sets.   Second, a noticeable way in which adding antidote data improves fairness is by reducing the magnitude of the outliers that drive unfairness in both training and testing.    Finally, the figure shows that in this example adding antidote data actually improves overall accuracy of the model predictions. 

Figure \ref{fig:group-example} shows the effect of optimal antidote data on per-group RMSE in the test set.    
For each group (genre) of movies, the corresponding point shows the group's RMSE before and after adding antidote data.   Additionally, the boxplots on each axis illustrate the distribution of RMSE values across groups before and after adding antidote data. 

First, we observe that all points are below the line $y=x$, i.e., adding antidote data improves the prediction accuracy of all genres and thus the overall accuracy of the model. Moreover, the boxplots show that improvements in rating estimations are so that the cross-group variability in RMSE is decreased to reach a fairer situation.
Finally, we see that outliers particularly benefit from addition of antidote data;  this can be seen as larger RMSE improvements in genres that initially had larger RMSE. In particular, \emph{Documentary} and \emph{Horror} have the largest prediction errors before adding antidote data, and their RMSEs are the most improved (furthest below $y=x$) after adding antidote data.
\section{Conclusion}
In this paper we propose a new strategy for improving the socially
relevant properties of a recommender system:  adding antidote data.  We
have presented an algorithmic framework for this strategy and applied it
to a range of socially important objectives.   Using this strategy, one does not need to modify the original system input data or the
system's algorithm.  We show that the resulting framework can
efficiently improve the polarization or fairness properties of a
recommender system.   We conclude that the developed framework can be a
flexible and effective approach to addressing the social impacts of a
recommender system.\\

\noindent\textbf{Acknowledgements.}
This research was supported by NSF grants CNS-1618207, IIS-1421759, and a European Research Council (ERC) Advanced Grant for the project ``Foundations for Fair Social Computing'' (grant no. 789373).

\appendix
\section{Derivation of the gradients}\label{sec:Gradients}
\subsection{Derivation of $\frac{\partial \matV}{\partial \tilde{\matX}}$}\label{sec:Gradients-model}
Let $E$ be the value of the objective function in (1). Assuming that the factorization algorithm finds a local optimum of $E$, we have $\frac{\partial E}{\partial \vecv_j} = 0$, which give us the following:

\begin{equation}
\sum_{i \in \Omega_j} (x_{ij} - \vecu_i^\intercal \vecv_j) \vecu_i +
\sum_{i=1}^{n'} (\tilde{x}_{ij} - \tilde{\vecu}_i^\intercal \vecv_j) \tilde{\vecu}_i = \lambda \vecv_j
\end{equation}
From the above equation we can show that the following formula for $\vecv_j$ holds at a local optimum of $E$:
\begin{equation}
\vecv_j =
(\sum_{i \in \Omega_j} \vecu_i \vecu^\intercal_i + \tilde{\matU} \tilde{\matU}^\intercal + \lambda \matI_{\ell})^{-1} (\sum_{i \in \Omega_j} x_{ij} \vecu_i + \sum_{i=1}^{n'} \tilde{x}_{ij} \tilde{\vecu}_i)
\end{equation}
Therefore, assuming that an infinitesimal change in $\tilde{x}_{ij}$ only results in first order corrections we   can write:
\begin{equation}
\frac{\partial \vecv_j}{\partial \tilde{x}_{ij}} =
(\sum_{i \in \Omega_j} \vecu_i \vecu^\intercal_i + \tilde{\matU} \tilde{\matU}^\intercal + 
\lambda \matI_{\ell})^{-1} \tilde{\vecu}_i
\end{equation}

\subsection{Gradients of the objective functions}\label{sec:Gradients-utilites}
\spara{Polarization}\\
\begin{align}
\frac{\partial R_{pol}}{\partial \hat{x}_{ij}} &=
\frac{2}{n^2 d} 
\sum_{k \neq i} (\hat{x}_{ij} - \hat{x}_{kj})\\
&= \frac{2}{nd} (\hat{x}_{ij} - \frac{1}{n} \sum_{k=1}^n \hat{x}_{kj})\\
&=\frac{2}{nd}( \hat{x}_{ij} - \mu_j)
\end{align}
where $\mu_j$ is the average estimated rating for item $j$.

\spara{Individual fairness}\\
For $(i,j) \in \Omega$ we have:
\begin{align}
\frac{\partial R_{indv}}{\partial \hat{x}_{ij}} &=\frac{1}{n^2}(
\sum_{k>i} 2(\ell_i - \ell_k) \frac{\partial \ell_i}{\partial \hat{x}_{ij}} +
\sum_{k<i} -2(\ell_k - \ell_i) \frac{\partial \ell_i}{\partial \hat{x}_{ij}}
)\\
&=\frac{2}{n^2}
\sum_{k \neq i} (\ell_i - \ell_k) \frac{\partial \ell_i}{\partial \hat{x}_{ij}} 
\\
&= \frac{4(\hat{x}_{ij} - x_{ij})}{n^2 {|\Omega^i|}} \sum_{k \neq i} (\ell_i - \ell_k)\\
&= \frac{4(\hat{x}_{ij} - x_{ij})}{n{|\Omega^i|}} (\ell_i - \frac{1}{n} \sum_{k=1}^n \ell_k )
\end{align}
therefore,

\begin{equation}
\begin{aligned}
&\frac{\partial R_{indv}}{\partial \hat{x}_{ij}} = 
\begin{cases}
    \frac{4(\hat{x}_{ij} - x_{ij})}{n {|\Omega^i|}} (\ell_i - \mu_{indv} )  & \quad (i,j) \in \Omega\\
    0      & \quad (i,j) \notin \Omega
\end{cases}
\end{aligned}
\end{equation}
where $\mu_{indv}$ is the average of user losses.

\spara{Group fairness}\\
Assume $\mathcal{G()}$ is a function that maps each user/item to its group label. For $(i,j) \in \Omega$ we have:
\begin{align}
\frac{\partial R_{grp}}{\partial \hat{x}_{ij}} &=\frac{2}{g^2}
\sum_{k \neq {\mathcal{G}(i)}} (L_{\mathcal{G}(i)} - L_k) \frac{\partial L_{\mathcal{G}(i)}}{\partial \hat{x}_{ij}} 
\\
&= \frac{4(\hat{x}_{ij} - x_{ij})}{g^2 {|\Omega_{\mathcal{G}(i)}|}} \sum_{k \neq {\mathcal{G}(i)}} (L_{\mathcal{G}(i)} - L_k)\\
&= \frac{4(\hat{x}_{ij} - x_{ij})}{g{|\Omega_{\mathcal{G}(i)}|}} (L_{\mathcal{G}(i)} - \frac{1}{g} \sum_{k=1}^g L_k )
\end{align}
therefore,

\begin{equation}
\begin{aligned}
&\frac{\partial R_{grp}}{\partial \hat{x}_{ij}} = 
\begin{cases}
    \frac{4(\hat{x}_{ij} - x_{ij})}{g {|\Omega_{\mathcal{G}(i)}|}} (L_{\mathcal{G}(i)} - \mu_G )  & \quad (i,j) \in \Omega\\
    0      & \quad (i,j) \notin \Omega\\
  \end{cases}                              
\end{aligned}
\end{equation}
where $\mu_G$ is the average of group losses.

\section{Heurisitc Algorithms}\label{sec:Fairness-heurestics}
In this section we present the pseudocode of the heuristic algorithms introduced in section 6.2 for generating antidote data to improve individual and group fairness.

\subsection{\texttt{heuristic1}}
\vspace{4pt}

\begin{enumerate}[label=\arabic*.]
    \item Start from a single antidote user $\tilde{\vecx}_1^{(0)}$.
    \item  Compute $\matU,\tilde{\vecu}_1,\matV = \Thetabold(\matX;\tilde{\vecx}_1^{(0)})$.
    \item Compute $\frac{\partial R(\hat{\matX})}{\partial \tilde{x}_{1j}}$ for each item $j$ using (9).
    \item If $\frac{\partial R(\hat{\matX})}{\partial \tilde{x}_{1j}} >0$ set $\tilde{x}_{1j} = \mathbb{M}_{min}$ Else set $\tilde{x}_{1j} = \mathbb{M}_{max}.$
    \item Copy $\tilde{\vecx}_1$ $\alpha$ times to generate $\tilde{\matX}$ for a given budget $\alpha.$ 
 \end{enumerate}

\subsection{\texttt{heuristic2}}
\vspace{4pt}
\begin{enumerate}[label=\arabic*.]
    \item Compute $\nabla_{\hat{\matX}} R(\hat{\matX})$ and reshape it into an $n \times d$ matrix $\matG$.
    \item Set $\vecd[j] = \vecg_j^\intercal \matU^\intercal \mathbf{1}_\ell$ for each item $j$ using $\matG$ and the original factor $\matU.$
    \item If $\vecd[j]>0$ set $\tilde{x}_{1j} = \mathbb{M}_{min}$ Else set $\tilde{x}_{1j} = \mathbb{M}_{max}.$
    \item Copy $\tilde{\vecx}_1$ $\alpha$ times to generate $\tilde{\matX}$ for a given budget $\alpha.$ 
 \end{enumerate}

\bibliographystyle{ACM-Reference-Format}
\bibliography{WSDM2019.bib}

\end{document}